\documentclass[nofootinbib,onecolumn]{revtex4}
\usepackage{graphicx}
\usepackage{subfig}
\usepackage{amssymb}
\usepackage{amsmath}
\usepackage{bm}
\usepackage{slashbox}
\begin{document}

\title{$f(R)$ global monopole revisited}
	
\author{Thiago R. P. Caram\^es$^1$}\email{thiago.carames@ufes.br}
\author{J\'ulio C. Fabris$^1$}\email{fabris@pq.cnpq.br}
\author{E. R. Bezerra de Mello$^2$}\email{emello@fisica.ufpb.br}
\author{H. Belich $^1$}\email{humberto.belich@ufes.br}

\affiliation{$^1$ Universidade Federal do Esp\'{\i}rito Santo (UFES), Av. Fernando Ferrari-514, 29075-910, Vit\'oria, ES - Brazil}
\vspace{1cm}
\affiliation{$^2$ Departamento de F\'isica, Universidade Federal da Para\'iba, 58.059-970, Caixa Postal 5.008, Jo\~ao Pessoa, PB - Brazil}


\begin{abstract} In this paper the $f(R)$ global monopole is reexamined. We provide an exact solution for the modified field equations in the presence of a global monopole for regions outside its core, generalizing previous results. Additionally, we discuss some particular cases obtained from this solution. We consider a setup consisting of a possible Schwarzschild black hole that absorbs the topological defect, giving rise to a static black hole endowed with a monopole's charge. Besides, we demonstrate how the asymptotic behavior of the Higgs field far from the monopole's core is shaped by a class of spacetime metrics which includes those ones analyzed here. In order to assess the gravitational properties of this system, we analyse the geodesic motion of both massive and massless test particles moving in the vicinity of such configuration. For the material particles we set the requirements they have to obey in order to experience stable orbits. On the other hand, for the photons we investigate how their trajectories are affected by the gravitational field of this black hole.

\textbf{Key-words}: Global monopole, $f(R)$ gravity.

PACS numbers: $04.50.+h$, $04.20.-q$
\end{abstract}


\maketitle

\section{Introduction}

One of the most important predictions expected within the grand unified theories (GUT) is the potential appearence of topological defects in the early universe. Such exotic configurations would arise due to the successive phase transitions experienced by the universe in its primordial stage. The spontaneous symmetry breaking (SSB) process triggered thanks to such phase transitions would left behind many kinds of topological defects, each one related
to the different types of symmetry group which would be broken down \cite{kibble,vilenkin}. The breaking of the global $SO(3)$ symmetry, for instance, gives rise to a spherically symmetric topological defect called global monopole. A simplified global monopole model was introduced in \cite{sokolov}. The gravitational effects of this object was investigated by Barriola and Vilenkin within the General Relativity (GR) framework \cite{barriola}. In the latter the authors have shown that the line element associated
with this defect corresponds to a solid deficit angle, what affects particularly the deflection of light rays moving near the monopole. The authors also raised a twofold interpretation for the mass term appearing in the solution for the corresponding Einstein field equations: the first possibility is assuming such term as the mass enclosed in the monopole's core, whereas the second one suggests that a star could have collapsed into a black hole near the global monopole resulting in a Schwarzschild black hole carrying the monopole's charge. 
The first case leads to a very tiny mass which is usually discarded as it is considered negligible at the astrophysical scale. This fact implies the non-existence of a newtonian potential generated by the global monopole which consequently prevents the monopole to capture massive test particles moving around it. Differently, light rays passing by the surroundings of the monopole would have their trajectories affected by the solid angle deficit, thus leaving behind a possible signature to detect the presence of the defect. 
Using the proper junction conditions for both the inner and outer regions of the defect's core, Hahari and Loust\`o have computed this tiny mass and found it to be negative \cite{lousto}. The direct physical interpretation for this negative mass is a repulsive gravitational potential that precludes stable orbits for timelike test particles moving in the global monopole spacetime.    

Although the well known success of the GR as an efficient description of the majority of the gravitational phenomena observed in the universe, some important challenges for the theory still persist.
The Einstein theory is plagued with singularities, seems incompatible with a quantum representation of the gravity and more recently faces apparent drawbacks
in the cosmological picture where it fails in offering a proper answer for the current accelerated expansion experienced by the cosmic background. The only way of addressing properly the observed cosmic speed-up within the GR scenario is by evoking a cosmological constant whose addition leads to inconsistencies between the cosmological observations and the quantum predictions for the vacuum density energy, the so-called cosmological constant problem.
Therefore, in the last decades it has been very common to see many proposals of modified theories of gravity, suggested as possible attemptions to provide an enlightenment to these issues. The $f(R)$ theories have been arisen as one of these possibilities. In the $80$'s these theories were suggested as a possible answer for the inflationary mechanism \cite{staro}. In the recent years such scenarios were extensively studied within the cosmological ambit
as a feasible way out for the dark energy issue, leading to strong constraints on the $f(R)$ models both at the background and the perturbative level. It has been shown that $f(R)$ theories possess an intimate connection with scalar-tensor theories with both the metric and Palatini $f(R)$ gravity representing two different versions of a Brans-Dicke gravity with a self-interaction potential: while the former is equivalent to the case in which the Brans-Dicke parameter is $\omega=0$, the latter corresponds to $\omega=-\frac{3}{2}$ \cite{soti,capo}. See \cite{sotiriou,odin,odin1} for detailed and comprehensive reviews on the $f(R)$ theories of gravity. 

The spacetime of the global monopole was already previously studied in the Brans-Dicke theory of gravity in \cite{romero} assuming the weak field approximation. The $f(R)$ gravity 
is another gravitational scenario in which the global monopole physics was also analysed \cite{carames1}. In both cases the authors made proper comparisons with the corresponding results 
obtained in the scope of GR. Moreover, their analysis also have in common the finding of the capability of the global monopole to trap test particles moving around it. Such outstanding feature 
was owing to the emergence of a newtonian potential and represents a prompt consequence of the new degrees of freedom coming from the modification of the gravity. Since the $f(R)$ global monopole was introduced in \cite{carames1}, it has received a lot of attention of some authors who studied several interesting physical phenomena within this context. The problem of a test particle moving in the surroundings of a global monopole was explored in a subsequent paper by these same authors \cite{carames2}, where some approximative assumptions were made. For example, the solutions for the corresponding field equations were obtained in the weak field regime and the departure from the GR was considered very small. Besides, the monopole was assumed as a point-like defect whose mass can be neglected at a astrophysical level. These same suppositions were admitted by \cite{man} where they studied the thermodynamics of the black hole provided with $f(R)$ global monopole charge. 
The $f(R)$ global monopole was also used to study strong lensing effects in \cite{cheng}, where the authors obtained analytic expressions for both the deflection angle and the time delay between multiple images in terms of the global monopole parameter and the $f(R)$ correction. In \cite{jpmg} the authors generalized the original solution by assigning rotation to the $f(R)$ global monopole. In \cite{valdir} this system has served as background to compute the quasinormal modes for scalar and spinor fields, by means of the WKB approximation. In \cite{melis} the authors claimed to have obtained an exact solution for the $f(R)$ global monopole. However, it seems that what they found, at the end of the day, was just a solution within standard GR, since they performed its analysis in a framework where $f(R)\propto R$ which is quite restrictive with respect to our study, where we do assume possible deviations from GR.  

In the present article we deepen the discussion about the $f(R)$ global monopole and add new contributions to the previous studies on this issue. In particular, we achieved an exact solution for the field equations outside the monopole's core for a specific class of $f(R)$ theories, enlarging the possibilities to be explored within such models. It is worth noting that the exact solution here obtained is an extension of what was uncovered by Multamaki and Vilja \cite{Mut} for the vacuum case, for the same class of $f(R)$ theories that we used here. We also demonstrated the form assumed by the Higgs field in the regions very far from the centre of the topological defect and showed how such asymptotic behavior is shaped by the background geometry given by a family of polynomial-like metric functions. This finding generalizes previous results obtained in the context of a global monopole in de Sitter/anti-de Sitter spacetime \cite{bertrand}. The next step was to investigate the geodesic motion of both massive and massless test particles. 
For timelike particles we carried out a detailed analysis of the necessary conditions for stable and circular orbits, by examining the properties of the corresponding effective potentials. On the other hand, the study of the orbit equation for light particles allowed us to determine the effects both of the deviation from GR and the global monopole's charge on the gravitational bending of light.  

In the next section we introduce the basic setup from where our analysis evolves and the standard global monopole model is briefly reviewed. In the section II, we provide an exact solution for the $f(R)$ global monopole and discuss some useful approximate cases where our studies are going to be developed. The consequences of the modification of the gravity theory on the
dynamics of massive test particles is assessed in the section III, where the expected requirements for stable orbits are properly appraised. In the section IV we study the light deflection issue in such gravitational field. 
Finally, the last section is dedicated to our concluding remarks and to the possible future investigations motivated by the present work.   

\section{The model}
\subsection{The $f(R)$ theory}
The action of the $f(R)$ gravity reads \cite{sotiriou} 
\begin{equation}
\label{action}
S=\frac{1}{2 \kappa} \int d^{4} x \sqrt{-g} f(R)+ {\cal S}_{m}, 
\end{equation}
where $\kappa=8\pi G$ and the matter action ${\cal S}_{m}$ is given in terms of the corresponding lagrangian density as follows
\begin{equation}
\label{Sm}
{\cal S}_{m}= \int d^{4} x \sqrt{-g} {\cal L}.
\end{equation}
In the metric formalism, the action (\ref{action}) is extremised with respect to the metric tensor providing the following field equations
\begin{equation}
\label{eqfR}
F(R)R_{\mu\nu}-\frac{1}{2}f(R)g_{\mu\nu}-\nabla_{\mu}\nabla_{\nu}F(R)+g_{\mu\nu}\Box F(R)=\kappa T_{\mu\nu}, 
\end{equation}
with $F(R)\equiv df(R)/dR$ and as usual $\mu,\nu=0,1,2,3$. As it is known, the energy-momentum tensor $T_{\mu\nu}$ is obtained from the lagrangian ${\cal L}$ as 
\begin{equation}
\label{emt0}
T_{\mu\nu}=\frac{2}{\sqrt{-g}}\frac{\delta \sqrt{-g}{\cal L}}{\delta g^{\mu\nu}}\ .
\end{equation}
Taking the trace of (\ref{eqfR}) we are left with  
\begin{equation}
\label{BoxF}
f(R)=\frac{1}{2}\left(F(R)R+3\Box F(R)-\kappa T\right).
\end{equation}
Let us recall that in GR the contracted Einstein field equations corresponds to an algebraic constraint involving the
Ricci scalar and the trace of the energy-momentum tensor, whereas in $f(R)$ theory contracting the field equations leads to a further dynamical equation for $F(R)$, which indicates that $F(R)$ plays the role of
an extra degree of freedom in the metric $f(R)$ gravity. Hence, (\ref{BoxF}) has an important meaning since it enhances a crucial distinction between GR and $f(R)$ theories. 

Using of the expression (\ref{BoxF}) we can also write the function $f(R)$ in terms of $F(R)$, its derivatives and the trace of the energy-momentum tensor. This allows us to get rid of $f(R)$ and promote $F(R)$ to the actual function to be specified in a given $f(R)$ theory.
It is possible to notice that such procedure reduces considerably the original complexity of the system of differential equations (\ref{eqfR}), making its integration much simpler. Furthermore, another interesting reason to work with $F(R)$ is its clear physical interpretation as a scalar
degree of freedom in $f(R)$ gravity. In principle it can make easier any possible comparison with results obtained within scalar-tensor theories of gravity. 
The purely radial dependence of the Ricci scalar, $R=R(r)$, leads to an alternative parametrization for $F(R)$, as $F(R(r))\equiv {\cal F}(r)$. Since in GR we have ${\cal F}(r)=1$, we can assume that any departure from the einsteinian theory appears in ${\cal F}(r)$ in the following way 
\begin{equation}
\label{F}
{\cal F}(r)=1+\psi(r), 
\end{equation}
where $\psi(r)$ is a function that encodes the modification of the gravity, whose functional form is arbitrary and must be specified in order to enables one to integrate the system of differential equations resulting from (\ref{eqfR}).   

\subsection{The global monopole spacetime} 
The model is described by the lagrangian density below \cite{barriola}
\begin{equation}
\label{lagr}
{\cal L}=\frac{1}{2}\partial_{\mu}\phi^{a} \partial^{\mu}\phi^{a}-\frac{1}{4} \lambda (\phi^{a}\phi^{a}-\eta^2)^2\ ,
\end{equation}
which exhibits the symmetry breaking of the $SO(3)$ to $U(1)$ groups. In the equation above the parameter $\lambda$ is a positive coupling constant and $\eta$ is the energy scale at which the symmetry is broken, The $SO(3)$-symmetric Higgs field $\phi^a$ is given by an isotriplet of scalar fields whose form corresponds to the well-known hedgehog {\it Ansatz}, 
\begin{equation}
\label{campo}
\phi^{a}=\eta h(r) \frac{x^{a}}{r}\ .
\end{equation}
With the index $a=1,2,3$ and $x^{a}x^{a}=r^2$. The radial function $h(r)$ is subject to the following boundary conditions 
\begin{equation}
\label{BCs}
h(0)=0,\;\;\;\; h(r\rightarrow \infty)=1.
\end{equation}
The spherically symmetric line element describing a spacetime around a static source can be written as 
\begin{equation}
\label{metric}
ds^2=B(r)dt^2-A(r)dr^2-r^2d\theta^2-r^2{\textrm{sin}^2\theta}d\varphi^2\ .
\end{equation}
Whereas the energy-momentum tensor for the global monopole, obtained from the Lagrangian (\ref{lagr}), has the following non-vanishing components read
\begin{eqnarray}
\label{emt}
T^{0}_{\;\;0}=\eta^2 \left[\frac{h'^2}{2A}+\frac{h^2}{r^2}+\frac{\lambda}{4}\eta^2\left(h^2-1\right)^2\right] ,\nonumber\\
T^{1}_{\;\;1}=\eta^2 \left[-\frac{h'^2}{2A}+\frac{h^2}{r^2}+\frac{\lambda}{4}\eta^2\left(h^2-1\right)^2\right] ,\\
T^{2}_{\;\;2}=T^{3}_{\;\;3}=\eta^2 \left[\frac{h'^2}{2A}+\frac{\lambda}{4}\eta^2\left(h^2-1\right)^2\right] ,\nonumber
\end{eqnarray}
where the prime denotes derivatives with respect to the radial coordinate $r$. The field equation for $\phi^{a}$ in the background (\ref{metric}) is
\begin{equation}
\label{phi}
\frac{h''}{A}+\left[\frac{2}{Ar}+\frac{1}{2B}\left(\frac{B}{A}\right)'\right]h'-\frac{2h}{r^2}-\lambda \eta^2 h\left(h^2-1\right)=0.
\end{equation}
In this model is assumed that far from the monopole's core, the function $h(r)$ tends to unity which leads the energy-momentum tensor to a very simple form given by
\begin{equation}
\label{emt1}
T_{\mu}^{\nu}\approx\textrm{diag}\left(\frac{\eta^2}{r^2},\frac{\eta^2}{r^2},0,0\right)\ .
\end{equation}
The gravitational implications due to a possible existence of global monopoles were first explored by M. Barriola and A. Vilenkin \cite{barriola}, who considered the approximation (\ref{emt1}) to achieve the solution for the gravitational field of this defect. They found that a spacetime associated with this object is characterized by a non-trivial topology observed as a deficit solid angle, which brings remarkable consequences to the light deflection phenomenon. On the other hand, neglecting the core's mass, they verified that the global monopole does not exert any gravitational force on the matter around it.    
Considering (\ref{metric}) and (\ref{emt}), they obtained the following solution for the Einstein field equations
\begin{equation}
\label{bv}
B(r)=A(r)^{-1}=1-8 \pi G \eta^2-2GM/r\ ,
\end{equation}
where $M$ is an integration constant that can be identified with the central mass that generates the gravitational field. For the solution (\ref{bv}), Barriola and Vilenkin discussed two possible physical interpretations for such mass term. The first one presents this parameter as the mass contained inside the radius of the monopole's core. In this case, this mass is usually neglected as
it is insignificant at astrophysical scales. The second hypothesis suggests that this term could be the mass of a Schwarzschild black hole that swallowed the global monopole and hence carries its charge. This latter case opens a door for a large variety of possible phenomena to be explored in the spacetime of such modified black hole.    

Global monopoles in the Brans-Dicke theory of gravity were also analyzed by A. Barros and C. Romero in \cite{romero}. Within the weak field regime they found an emerging force which acts on the test particles moving in the monopole's spacetime. In \cite{carames1} the weak field approximation was also considered to study this cosmic defect in the context of a $f(R)$ theory.
In that paper the authors purchased results very similar to the Brans-Dicke case, at least from a qualitative point of view, as for instance observing the appearence of an extra force that accounts for the capture of test particles moving around the monopole. Such feature is absent in the general relativistic scenario and can be assigned to the influence of the scalar degree
of freedom which arises in the both theories and represents an evident heritage of the modification of the gravity.       

\section{Solutions for the field equations}
To obtain the explicit form of the field equations for this model we follow the same procedure adopted in \cite{carames1,Mut}. Substituting (\ref{BoxF}) into (\ref{eqfR}) the resulting equation can be written as
\begin{equation}
\label{eq0}
F(R)R_{\mu\nu}-\nabla_{\mu}\nabla_{\nu}F(R)-\kappa T_{\mu\nu}=\frac{g_{\mu\nu}}{4}\left[F(R)R-\Box F(R)-\kappa T\right]. 
\end{equation}
This expression tell us that the following combination, involving rank-$2$, diagonal, tensor quantities 
\begin{equation}
\label{comb}
C_{\mu}\equiv \frac{F(R)R_{\mu\mu}-\nabla_{\mu}\nabla_{\mu}F(R)-\kappa T_{\mu\mu}}{g_{\mu\mu}},
\end{equation}
whose indices $\mu$ are fixed\footnote{This means there is no summation here.}, does not depend on the mentioned index, so that the relation below
\begin{equation}
\label{Cmu}
C_{\mu}-C_{\nu}=0,
\end{equation}
keeps valid for any $\mu$ and $\nu$. One can check that (\ref{Cmu}) allows for the obtaining of the only two independent field equations, corresponding to $C_{1}-C_{0}=0$ and $C_{2}-C_{0}=0$.
So, these two relations will provide a system of differential equations involving $A(r)$, $B(r)$ and ${\cal F}(r)$ for a given matter field. As said before, for the function ${\cal F}(r)$ a suitable {\it Ansatz} has to be adopted in order to determine the solution of the system. For a static, spherically symmetric matter configuration the following field equations hold:
\begin{equation}
\label{eq1}
2r{\cal F}''-2{\cal F}\frac{Y'}{Y}-r{\cal F}'\frac{Y'}{Y}=16 \pi G (T_{1}^{1}-T_{0}^{0})A r\ ,
\end{equation}
and
\begin{equation}
\label{eq2}
-4B+4Y-4rB\frac{{\cal F}'}{{\cal F}}+2r^2B'\frac{{\cal F}'}{{\cal F}}+2rB\frac{Y'}{Y}-r^2B'\frac{Y'}{Y}+2r^2B''+\frac{32 \pi G}{\cal F}Y r^2 (T_{2}^{2}-T^{0}_{0})=0\ ,
\end{equation}
where $Y(r)\equiv A(r)B(r)$. For the global monopole energy-momentum tensor (\ref{emt}) these equations are rewritten as follows
\begin{equation}
\label{eq1}
2r{\cal F}''-2{\cal F}\frac{Y'}{Y}-r{\cal F}'\frac{Y'}{Y}=-16 \pi G \eta^2 h'^2 r\ ,
\end{equation}
and
\begin{equation}
\label{eq2}
-4B+4Y-4rB\frac{{\cal F}'}{{\cal F}}+2r^2B'\frac{{\cal F}'}{{\cal F}}+2rB\frac{Y'}{Y}-r^2B'\frac{Y'}{Y}+2r^2B''-\frac{32 \pi G \eta^2 h^2 Y}{\cal F}=0\ .
\end{equation}
Once a convenient shape for ${\cal F}(r)$ is assumed, the equations (\ref{eq1}) and (\ref{eq2}) along with (\ref{phi}) form a system of equations whose solution can be in principle obtained numerically given a proper set of boundary conditions. It is common to consider the outside the core approximation, for which $h\approx1$ and the global monopole energy-momentum tensor is given by (\ref{emt1}), which makes the form of the dynamical equations even simpler \cite{carames1} 
\begin{equation}
\label{EQ1}
2r{\cal F}''-2{\cal F}\frac{Y'}{Y}-r{\cal F}'\frac{Y'}{Y}=0\ ,
\end{equation}
\begin{eqnarray}
\label{EQ2}
-4B+4Y-4rB\frac{{\cal F}'}{{\cal F}}+2r^2B'\frac{{\cal F}'}{{\cal F}}+2rB\frac{Y'}{Y}-r^2B'\frac{Y'}{Y}+2r^2B''-\frac{32 \pi G \eta^2 Y}{\cal F}=0\ .
\end{eqnarray}
Let us notice that the solution for the system of equations above can only be achieved if the shape of ${\cal F}$ is specified beforehand. 
As mentioned before we are interested in functions ${\cal F}$ with the form (\ref{F}) and that means, at the end of the day, to specify the expression for the deviation from GR, $\psi(r)$. For convenience, we adopt the simplest {\it Ansatz} for such function and assume it as linear in the radial coordinate, $\psi(r)=\psi_0 r$. This is a usual choice considered in the literature.    

For ${\cal F}=1+\psi_0 r$, (\ref{EQ1}) implies that $Y(r)=Y_0=$const., simplifying enormously the equation (\ref{EQ2}) which now gets reduced to an ordinary second-order differential equation for $B(r)$, whose integration gives
\begin{eqnarray}
\label{sol}
&&B(r)=Y_{0}(1-8\pi G \eta^2)+\frac{c_1 \psi_0}{2}-\frac{c_1}{3 r}-r \psi_0 \left[Y_0(1-16 \pi G \eta^2)+\psi_0 c_1\right]\nonumber\\&+&\frac{r^2}{2}\left\{\psi_0^2Y_0(3-32 \pi G \eta^2)+2c_2+2\psi_0^2\left[Y_0(1-16\pi G \eta^2)+\psi_0 c_1\right] \ln \left(\psi_0+\frac{1}{r}\right)\right\}\ ,
\end{eqnarray}
where $c_1$ and $c_2$ are integration constants. Comparing (\ref{sol}) with the Eq. (20) of \cite{Mut}, the solution above can easily be seen as a clear generalization of that vacuum result for the case in which a global monopole sources the gravitational field. 
Besides, notice that this solution is more general than that ones found in previous studies as it carries corrections that are absent in all those approaches, where the approximations $\left|\psi_0 r\right|\ll1$ and the weak field limit were taken into account \cite{carames1}.
It is easy to verify such statement by looking at each term of (\ref{sol}) and comparing with the Eq. (36) of \cite{carames1}. In order to help us in such comparison, let us fix $c_2=0$. Assuming the smallness of the correction on GR we can keep just the linear terms in $\psi_0 r$ and throw away all the higher order contributions. 
This enables us to neglect all the terms between the curly brackets in (\ref{sol}). The integration constant $c_1$ is usually associated with the newtonian potential, by means of the identification $c_1=6GM$. In the approximation adopted in \cite{carames1} this mass term is a first order contribution, as well as the monopole's charge and the modification of gravity parameter, $G\eta^2$ and $\psi_0$, respectively.
Therefore, if the correction on GR is tiny and the weak field approximation applies it is reasonable to assume that all the crossing terms involving $GM$, $\psi_0$ and $G \eta^2$ can be ruled out\footnote{In fact it is known that tipically we have $G\eta^2\sim 10^{-6}$ \cite{vilenkin}.}, so that just the purely linear terms in each one of these quantities survive. Lastly, by setting $Y_0=1$ the expression (\ref{sol}) takes the form  
\begin{equation}
\label{solB}
B(r)=1-8\pi G \eta^2-\frac{2GM}{r}-\psi_0 r\, 
\end{equation}
which corresponds to the original solution found by Caram\^es {\it et al} \cite{carames1}. So, we have just shown that (\ref{solB}) is just a particular case of the solution (\ref{sol}) obtained by us in the present work. The metric given by (\ref{solB}) is quite used to address the $f(R)$ global monopole ploblem and its possible applications, as can be seen in our references.
This suggests that the solution presented here can bring possible corrections to the studies previously performed involving an $f(R)$ global monopole.

For convenience let us rewrite (\ref{sol}) as
\begin{eqnarray}
\label{sol1}
B(r)=\beta-\frac{c_1}{3 r}-(2\beta-Y_0) \psi_0 r+\frac{r^2}{2}\left[\psi_0^2Y_0(3-32 \pi G \eta^2)+2c_2+2\psi_0^2 (2\beta-Y_0) \ln \left(\psi_0+\frac{1}{r}\right)\right]\ ,
\end{eqnarray}
where $\beta$ denotes the combination $\beta=Y_0(1-8\pi G \eta^2)+\frac{c_1 \psi_0}{2}$. Now let us analyse some particular cases arising from the opportune choice of the integration constants present in this solution. The first case refers to an exact solution, whereas the second one takes some useful approximations into account.

\subsection{For $Y_0=2\beta$}
We can discard the logarithmic term by imposing $Y_0=2\beta$, which consequently implies in setting also the $r$-term to zero. Additionally, we can choose the $r$-independent term as $\beta=1-8 \pi G \eta^2$, which allows for the prompt recovering of the standard Barriola-Vilenkin solution in the limits $\psi_0\rightarrow 0$ and $Y_0\rightarrow 1$, with the integration constant $c_1=6GM$.
Consequently, these choices fix the forms both for $Y_0$  
\begin{equation}
\label{Y0}
Y_0=2(1-8 \pi G \eta^2),
\end{equation}
and for the modification of gravity parameter
\begin{equation}
\label{psi0}
\psi_0= \frac{-(1-8 \pi G \eta^2)(1-16 \pi G \eta^2)}{3GM}.
\end{equation}
Notice that in the absence of global monopole, $\eta=0$, the variables $Y_0$ and $\psi_0$ reduce to those ones found by Multamaki and Vilja in \cite{Mut} (see the ``solution I'' obtained by these authors) within the vacuum context. Interestingly, as can be seen in (\ref{psi0}) the solution considered here imposes a constraint on the value of the parameter $\psi_0$, by fixing it in terms of the numerical values both of the source mass and the monopole's charge, once these quantities are known.
Taking all these assumptions into account the solution (\ref{sol}) can be written as  
\begin{equation}
\label{sol2}
B(r)=1-\frac{2GM}{r}-8 \pi G \eta^2-\frac{\tilde{\Lambda}}{3}r^2\ ,
\end{equation}
where we require that $c_2=-\frac{\Lambda}{3}$ in order to have a Schwarzschild-de Sitter (SdS) solution in the suitable limit. Moreover, $\tilde{\Lambda}$ plays the role of an effective cosmological constant given by $\tilde{\Lambda}\equiv\Lambda-3\psi_0^2(1-8\pi G \eta^2)(3-32 \pi G \eta^2)$ which displays the effects of the $f(R)$ correction on the usual cosmological constant, $\Lambda$. For a small correction around GR it is reasonable to neglect this extra term. So, the line element for this solution shall be
\begin{equation}
\label{line}
ds^2=\left(1-\frac{2GM}{r}-8 \pi G \eta^2-\frac{\tilde{\Lambda}}{3}r^2\right)dt^2-2(1-8 \pi G \eta^2) \left(1-\frac{2GM}{r}-8 \pi G \eta^2-\frac{\tilde{\Lambda}}{3}r^2 \right)^{-1}dr^2-r^2d\Omega^2\ ,
\end{equation}
where $d\Omega^2=d\theta^2+{\textrm{sin}^2\theta}d\varphi^2$. 

\subsection{For $Y_0=1$}
We can otherwise set $Y_0=1$ from the beginning and keep $\beta=Y_0(1-8\pi G \eta^2)+\frac{c_1 \psi_0}{2}$. This means to let the parameter $\psi_0$ free, instead of restricting it so strongly as was done in the previous case. Moreover, assuming a diminute deviation from GR the constant $\psi_0$ can be considered very tiny,
what enables one to retain just the linear powers of such parameter in the general solution (\ref{sol1}), so that we have
\begin{equation}
\label{sol3}
B(r)=1-8 \pi G \eta^2+3GM\psi_0-\frac{2GM}{r}-\psi_0 r\ . 
\end{equation}
Notice that in this approximative treatment a possible contribution of the independent term $\frac{c_1 \psi_0}{2}$ is taken into account, differently what is verified in the weak field approximation where such term does not come up, as can be seen in (\ref{solB}).
In the case in which the mass is interpreted as that one enclosed in the monopole's core, such term could be ignored at astrophysical level and the both solutions (\ref{solB}) and (\ref{sol3}) would match each other, coinciding with that analysed in \cite{carames1}. 
On the other hand, if this mass is due to a black hole that swallowed the monopole its contribution can be relevant. Its immediate effect seems to be either increasing or decreasing the effective magnitude of the monopole's charge, $8 \pi G \eta^2$, depending on whether the $\psi_0$'s sign is negative or positive, respectively. This extra term may exert a detectable influence on the motion of test massive or massless particles in the spacetime of this black hole.             

Interestingly, for $8 \pi G \eta^2=0$, the solution (\ref{sol3}) resembles to the static and sphericall symmetric solution obtained by P. D. Mannhein and D. Kazanas within a vacuum Weyl conformal gravity \cite{mann}.
\subsection{Asymptotic behavior of $h(r)$}
 Considering appropriate boundary conditions for the metric components $B(r)$ and $A(r)$ as well as the corresponding ones satisfied by $h(r)$ (\ref{BCs}), the equation (\ref{phi}) can be in principle solved through a numerical approach. However, we will restrict our study to the asymptotic case $h\rightarrow 1$ at infinity by analysing how one expects the solution behaves at this regime. In this vein, it is useful to assume a series expansion for the radial function $h(r)$ as follows 
\begin{equation}
\label{hAsy}
h(r)=1+\sum a_{n} r^{-n},
\end{equation}
where $n\geq 1$. The special case $A(r)B(r)=Y_0$ leads (\ref{phi}) to a simpler form
\begin{equation}
\label{phi1}
\frac{h''B}{Y_0}+\left(\frac{2B}{Y_0r}+\frac{B'}{Y_0}\right)h'-\frac{2h}{r^2}-\lambda \eta^2 h\left(h^2-1\right)=0.
\end{equation}
For sake of simplicity let us write the function $B(r)$ as
\begin{equation}
\label{B1}
B(r)=B_0+\frac{B_1}{r}+B_2 r+B_3 r^2.
\end{equation}
Let us notice that the solutions given by (\ref{solB}), (\ref{sol2}) and (\ref{sol3}) can be expressed in the form (\ref{B1}). Using (\ref{hAsy}) and (\ref{B1}) in (\ref{phi1}) we can determine the coefficients $a_n$ and then obtain the expression for $h(r)$ in the regions very far from the monopole's core.
\begin{eqnarray}
\label{hAsy1}
&&h(r\rightarrow \infty)=1-\frac{Y_0}{(B_3+\lambda \eta^2 Y_0)}\frac{1}{r^2}+\frac{Y_0\left[3Y_0^2\lambda \eta^2+2(B_0-Y_0)(B_3+\lambda \eta^2 Y_0)\right]}{2(B_3+\lambda \eta^2 Y_0)^2(2B_3-\lambda \eta^2 Y_0)}\frac{1}{r^4}+\nonumber\\+&&\frac{\left\{8Y_0B_1(B_3+\lambda \eta^2 Y_0)^2(2B_3-\lambda \eta^2 Y_0)-8B_2 Y_0\left[3 Y_0^2 \lambda \eta^2+2(B_0-Y_0)(B_3+\lambda \eta^2)\right]\right\}}{2(B_3+\lambda \eta^2 Y_0)^2(2B_3-\lambda \eta^2 Y_0)(10B_3-2 \lambda \eta^2 Y_0)}\frac{1}{r^5}+O\left(\frac{1}{r^6}\right),
\end{eqnarray}
whose definite form depends upon the parameters $B_0$, $B_1$, $B_2$, $B_3$ and $Y_0$ which characterize the different solutions. We provide the table below (Table 1) in order to help the reader to obtain promptly the explicit form for the asymptotic behavior of $h(r)$ for different metric tensors subject to the parametrization (\ref{B1}), in particular those ones mentioned or obtained throughout this paper. It is easy to verify that such general expression (\ref{hAsy1}) recovers the result presented in \cite{bertrand}, in the context of global monopoles within de Sitter/anti-de Sitter spacetimes \footnote{This can be checked by comparing (\ref{hAsy1}) with the eq. (11) derived by Bertrand {\it et al}.}.
\begin{table}[h]
\centering
\label{TAB}
\begin{tabular}{|c|c|c|c|c|c|}
\hline
\backslashbox{Solutions}{$B(r)$-Parameters} & $B_0$ & $B_1$ & $B_2$ & $B_3$ & $Y_0$ \\ \hline
(i) &    $1-8\pi G \eta^2$   & $-2GM$      &   $-\psi_0$    &    $0$   &    $1$   \\ \hline
(ii) &    $1-8\pi G \eta^2$  &    $-2GM$   &    $0$   & $-\tilde{\Lambda}/3$     &  $2(1-8 \pi G\eta^2)$     \\ \hline
(iii) &    $1-8\pi G \eta^2+3GM\psi_0$   &   $-2GM$    &   $-\psi_0$    &   $0$    &   $1$    \\ \hline
Barriola-Vilenkin & $1-8\pi G \eta^2$ & $-2GM$ & $0$ & $0$ & $1$ \\ \hline
\end{tabular}
\caption{Corresponding values of the constants $B_0$, $B_1$, $B_2$, $B_3$ and $Y_0$ for the solutions (\ref{solB}), (\ref{sol2}) and (\ref{sol3}). For convenience, in the table they are denoted merely as (i), (ii) and (iii), respectively. Additionally, we include the standard Barriola-Vilenkin solution.}
\end{table}
\section{Test particles around a black hole with an $f(R)$ global monopole}
It is well known that the geodesic motion of test particles in a certain spacetime obeys a Lagrangian, ${\cal L}_g$, given by
\begin{equation}
\label{Lg}
{\cal L}_g=g_{\mu\nu} \frac{d x^{\mu}}{d \tau} \frac{d x^{\nu}}{d \tau}=\epsilon, 
\end{equation}
where $\epsilon=0$ or $1$ labels massless and massive particles, respectively. The affine parameter, $\tau$, represents the proper time for massive particles that follow timelike geodesics.
Considering that the motion lies on the equatorial plane $\theta=\frac{\pi}{2}$, the Lagrangian ${\cal L}_g$ obtained from (\ref{metric}) gets reduced to   
\begin{equation}
\label{Lg1}
{\cal L}_g= B(r)\left(\frac{dt}{d\tau}\right)^2-A(r)\left(\frac{dr}{d\tau}\right)^2-r^2 \left(\frac{d\varphi}{d\tau}\right)^2,
\end{equation}
with $\tau$ being the proper time. Notice that the coordinates $t$ and $\varphi$ are cyclic implying in the following conserved quantities
\begin{equation}
\label{ctes}
E\equiv B(r) \frac{d t}{d\tau},\;\;\;\;\;\;\;\;L\equiv r^2 \frac{d \varphi}{d\tau},
\end{equation}
where $E$ ($L$) means the total energy (angular momentum) at infinity per unit particle rest mass. Using them we can express (\ref{Lg1}) as 
\begin{equation}
\label{en}
\frac{1}{2}\left(\frac{d r}{d \tau}\right)^2+V_{\textrm{eff}}(r)={\cal E},
\end{equation}
where ${\cal E}\equiv \frac{E^2}{2Y_0}$ and the effective potential reads
\begin{equation}
\label{veff}
V_{\textrm{eff}}(r)\equiv\frac{B(r)}{2Y_0}\left(\frac{L^2}{r^2}+1\right),  
\end{equation}
where we have set $\epsilon=1$. Equation (\ref{en}) reveals that the existence of the motion of a test particle is subject to the condition ${\cal E}-V_{\textrm{eff}}(r)>0$.
Besides, the values of $r$ for which ${\cal E}=V_{\textrm{eff}}(r)$ correspond to the turning points of the motion. The equations (\ref{en}) and (\ref{veff}) above are useful to investigate such possible orbits to be experienced either by massive or massless particles moving around a central black-hole. 
For instance, the GR shows us that considering a Schwarzschild background, for the geodesic motion of massive particle there is a minimum radius at which stable circular orbits are possible. It is denoted by innermost stable circular orbit (ISCO) and corresponds to   
$r_{\textrm{\tiny ISCO}}=6GM$ \cite{kaplan,landau}. The angular momentum of the particle for which such condition is achieved is $L_{\textrm{\tiny ISCO}}=2\sqrt{3}GM$ and the respective total energy is also promptly obtained $E_{\textrm{\tiny ISCO}}=\sqrt{8/9}$. 
If the Schwarzschild black hole carries a global monopole charge the ISCO parameters are slightly modified by scaling $M\rightarrow M/(1-8 \pi G \eta^2)$ in both $r_{\textrm{\tiny ISCO}}$ and $L_{\textrm{\tiny ISCO}}$, whereas $E$ becomes $E=\sqrt{(1-8 \pi G \eta^2)8/9}$ \cite{dadhich}. So let us use these scaled ISCO parameters
as reference in order to assess how they are influenced by the modification of the gravity. Before to proceed with our numerical treatment let us parametrize the main phyisical quantities used in our investigation into new dimensionless variables:
\begin{equation}
\label{dim0}
x\equiv r/GM; \;\;\;\; l\equiv L/GM\;\textrm{and}\;\;\;\sigma\equiv\psi_0 GM.
\end{equation}
For convenience, the results of this section will be expressed in terms of such variables. In this section, our interest is basically to analyse how far it is possible to deviate from GR without preventing the existence of circular stable orbits. It is expected that any stable circular motion satisfies the following requirements:
\begin{itemize}
 \item $\dot{r}=0$;
 \item $\partial V_{\textrm{eff}}/\partial r=0$;
 \item $\partial^2 V_{\textrm{eff}}/\partial r^2>0$,
\end{itemize}
where the dot denotes derivative with respect to the proper time. The second condition provides a polynomial equation whose roots give the radial positions where the effective potential have critical points, enabling the existence of orbital motion at those radii. The condition $\dot{r}=0$ results in $V_{\textrm{eff}}={\cal E}$, which shall fix the energy a particle orbiting the central mass at a certain radius $r$ (any of the roots of $\partial V_{\textrm{eff}}/\partial r=0$) should have in order to undergo a stable orbit. The sign of $\partial^2 V_{\textrm{eff}}/\partial r^2$ evaluated at the radial distance where a given orbit lies, tell us if these extrema of the effective potential correspond to minimum (positive sign) or maximum points (negative sign), 
resulting in stable or unstable orbits, respectively. A stable orbit means that the particle will tend to return immediately to its original orbit after being slightly disturbed. On the other hand, a particle undergoing an unstable orbital motion departs further from its original orbit when a slight flick is applied on it.

Let us now implement this procedure for some of the solutions discussed in this paper.  
\subsection{Case $B(r)=1-8\pi G \eta^2-\frac{2GM}{r}-\psi_0 r$}
In terms of (\ref{dim0}) this solution is written as 
\begin{equation}
\label{solX}
B(x)=1-\alpha^2-\frac{2}{x}-\sigma x,
\end{equation}
where we defined $\alpha^2\equiv 8\pi G \eta^2$. The algebraic equation $B(x)=0$ gives us two real roots, which correspond to the two horizons of this solution. Namely, these are an event and cosmological ones given by
\begin{equation}
\label{xH1}
x_{\textrm{h},1}=\frac{1-\alpha^2-\sqrt{(1-\alpha^2)^2-8 \sigma}}{2\sigma},
\end{equation}
and
\begin{equation}
\label{xH1}
x_{\textrm{h},2}=\frac{1-\alpha^2+\sqrt{(1-\alpha^2)^2-8 \sigma}}{2\sigma},
\end{equation}
respectively. Besides, (\ref{solX}) leads the effective potential (\ref{veff}) to take the form
\begin{equation}
\label{veffx}
V_{\textrm{eff}}(x)=\frac{(1-\alpha^2)}{2}-\left(\frac{\sigma l^2}{2}+1\right)\frac{1}{x}+\frac{l^2\left(1-\alpha^2\right)}{2x^2}-\frac{l^2}{x^3}-\frac{\sigma x}{2}.
\end{equation}
In Fig. $1$ we plot the effective potential (\ref{veffx}) against $x$, where we have set $l^2=12/(1-\alpha^2)^2$ and, by way of illustration, $\alpha^2=0.01$. The blue curve displays the Barriola-Vilenkin case, where the only difference with respect to the purely Schwarzschild black hole situation (without any global monopole inside it) is that the respective $x_{\textrm{\tiny ISCO}}$ 
gets increased by a factor $1/(1-\alpha^2)$ as we discussed before. So, in the $\sigma=0$ case the corresponding curve shows a minimum at $x_{\textrm{\tiny ISCO}}\approx 6.06$. For $\sigma=10^{-3}$, the radius of ISCO is slightly enhanced, giving $x_{\textrm{\tiny ISCO}}\approx 7.014$. Additionally, a local maximum point is verified at $x_{\textrm{\tiny max},1}\approx 5.53$, revealing a new feature in comparison with $\sigma=0$: even for a small departure from GR unstable orbits can also exist, besides the stable one. A second maximum point appears at $x_{\textrm{\tiny max},2}\approx 37.69$, as it is more clearly illustrated in the Fig. $1b$. It is easy to check that $V_{\textrm{\tiny eff}}(x_{\textrm{\tiny max},1})<V_{\textrm{\tiny eff}}(x_{\textrm{\tiny max},2})$, so a particle coming from very far regions has to possess energy greater than $V_{\textrm{\tiny eff}}(x_{\textrm{\tiny max},2})$ in order to cross the inner horizon and fall into the black hole. If such energy is smaller than this value, the test particle will approach the black hole at most up to the turning point, $E^2=V_{\textrm{\tiny eff}}(x_{\textrm{\tiny max},1})$, 
where it shall reverse its motion and proceed towards the infinity. 
\begin{figure}[!tbp]
  \centering
  \subfloat[]{\includegraphics[width=0.5\textwidth]{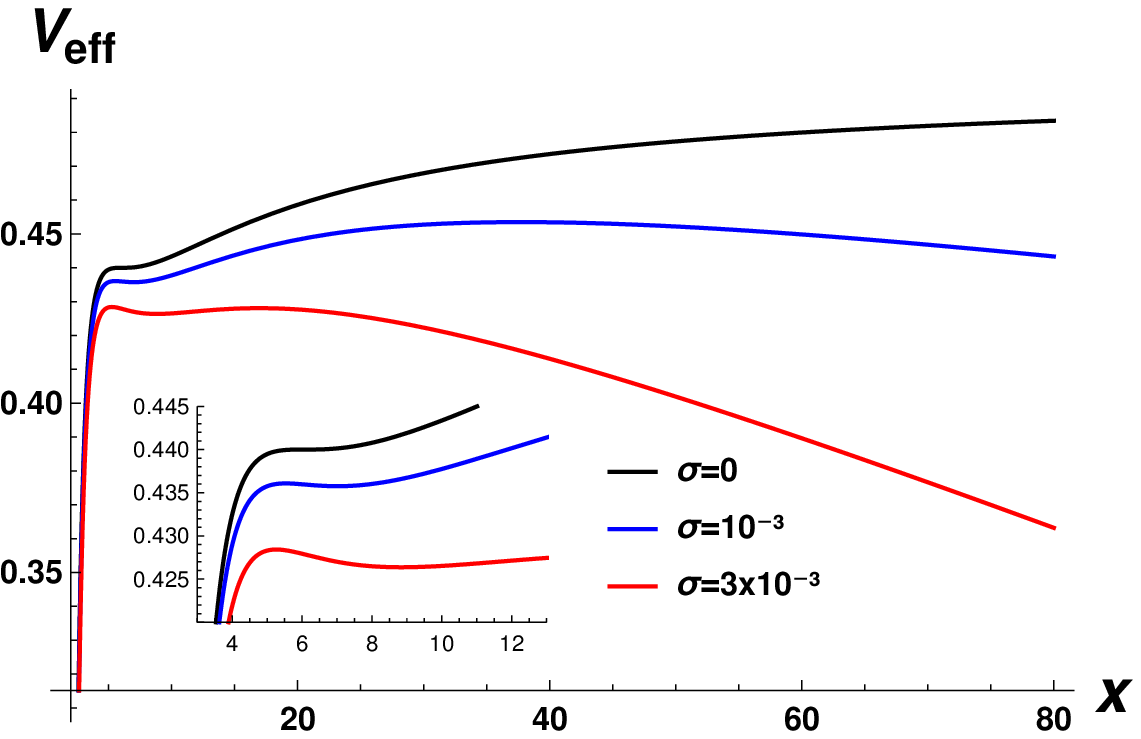}\label{fig:f1}}
  \hfill
  \subfloat[]{\includegraphics[width=0.5\textwidth]{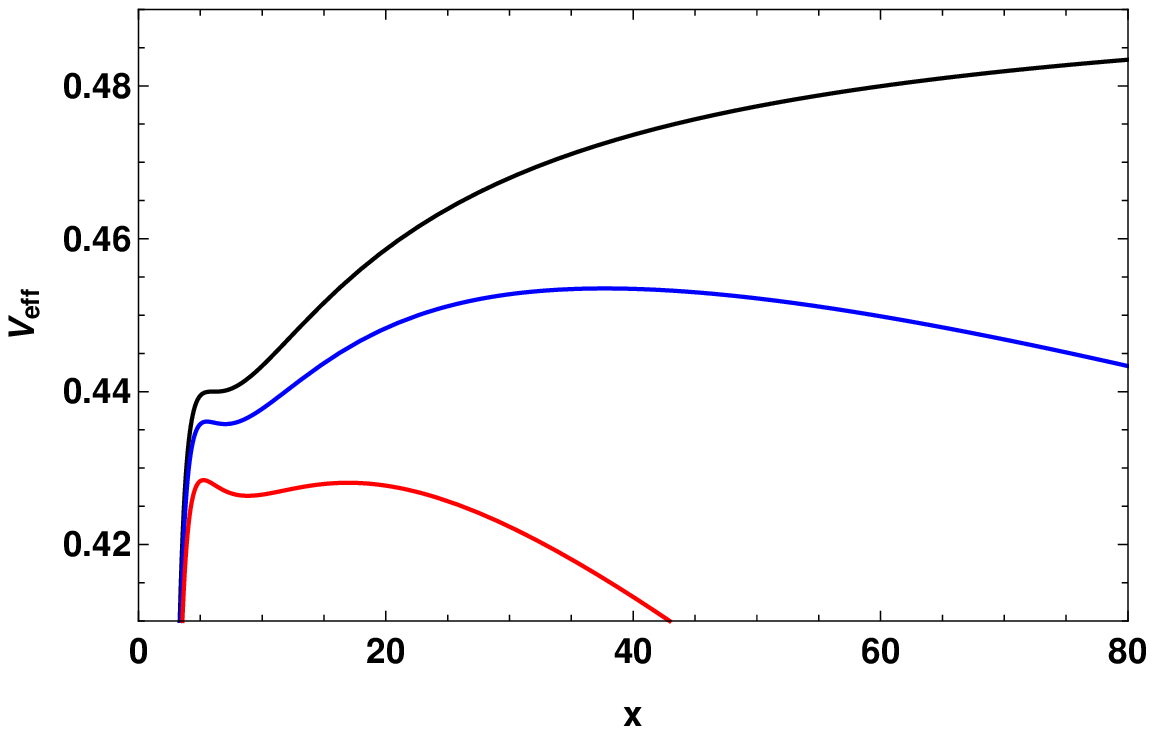}\label{fig:f2}}
  \caption{Variation of the effective potential with $x$. Here we fix $l^2=\frac{12}{(1-\alpha^2)^2}$ and $\alpha^2=0.01$. The corresponding ISCO position gets farther and farther from the central black hole, as the deviation from GR increases. This becomes a bit more evident by looking at the smaller graph on the panel (a), where the minima of the effective potential moves slightly to the right for incresing values of $\sigma$. Moreover, the modification of the gravity gives rise to maximum points for the $V_{\textrm{eff}}$ curve. The greater are the $\sigma$ values, the closer such points are to the event horizon. The panel (b) illustrates more clearly the appearence of these maximum points in the both blue and red curves, indicating that extra unstable orbits can also exist along with the stable ones.}
\end{figure}

For $\sigma=3\times10^{-3}$ we verified that the ISCO lies a bit farther from the central mass, at the position $x_{\textrm{\tiny ISCO}}\approx 8.84$. Besides, two maximum points for $V_{\textrm{\tiny eff}}$ shows up at $x_{\textrm{\tiny max},3}\approx 5.26$ and $x_{\textrm{\tiny max},4}\approx 16.96$, yielding a profile quite similar to the case $\sigma=10^{-3}$ from the qualitative point of view. So, it could be possible to have a stable circular motion encompassing an inner unstable orbit while is encircled by another one. However, differently from the case $\sigma=10^{-3}$, the outer unstable orbit gives for the effective potential a smaller value when comparing with that one given by the inner unstable orbit, i.e $V_{\textrm{\tiny eff}}(x_{\textrm{\tiny max},3})>V_{\textrm{\tiny eff}}(x_{\textrm{\tiny max},4})$, which means that any particle endowed with energy ${\cal E}>V_{\textrm{\tiny eff}}(x_{\textrm{\tiny max},3})$ is able to overcome the effective potential barrier and reach the centre of the black hole. When $\sigma>3\times10^{-3}$ we observed that only maximum points show up, indicating that stable orbits are not allowed for such cases.

One can verify that the solution (\ref{sol3}) yields an effective potential whose profile is quite similar to (\ref{veffx}), so that it exhibits the same general properties that we have approached in this subsection. The main difference appears in the displacement of the extrema of $V_{\textrm{eff}}$ when comparing with the case where $3GM\psi_0$ is absent. Nonetheless, we have checked that such discrepancies do not overcome $10\%$ of the corresponding values obtained without taking such extra term into account. Moreover, we noticed that the upper limit mentioned above, $\sigma=3\times 10^{-3}$, to be obeyed by (\ref{veffx}) in order to allow for stable orbits keeps valid even for this case. So, these features suggest that $3GM\psi_0$ does not add any significant contribution to the geodesic motion of material particles. 

\subsection{Case $B(r)=1-\frac{2GM}{r}-8 \pi G \eta^2-\frac{\tilde{\Lambda}}{3}r^2$}
For convenience, let us consider $\Lambda=0$ in the solution (\ref{sol2}), so that all the effects of the effective cosmological constant $\tilde{\Lambda}$ is due only to the deviation from GR. This assumption could be implemented from the beginning by merely defining the integration constant $c_2=0$. Such choice does not seem so weird if we recall that one of the main motivations for the $f(R)$ theories is exactly to dispense the cosmological constant, by replacing it by extra degrees of freedom of geometric nature which, in principle, would be able to play the same role that $\Lambda$ has in the dynamics of the current universe. Thus, (\ref{sol2}) becomes
\begin{equation}
\label{sol4}
B(r)=1-\frac{2GM}{r}-8 \pi G \eta^2+\psi_0^2(1-8\pi G \eta^2)(3-32\pi G \eta^2)r^2.
\end{equation}
As remarked previously, the solution above arises after a convenient choice for the integration constants of (\ref{sol}) which allowed for a vanishing of that undesirable logarithmic term. The price one had to pay for achieving such a simpler version of the exact solution was to find a non-trivial constraint relation, (\ref{psi0}), involving the parameters $\psi_0$, $8\pi G \eta^2$ and $M$, which prevents $\psi_0$ to vary freely by conveying to this parameter a tight dependence on the numerical values of both the mass parameter and the monopole's charge. Now let us employ (\ref{psi0}) and (\ref{dim0}) to express (\ref{sol4}) as follows
\begin{equation}
\label{solX1}
B(x)=1-\frac{2}{x}-\alpha^2+\frac{1}{9}(1-\alpha^2)^3(1-2\alpha^2)^2(3-4\alpha^2)x^2,
\end{equation}
whose corresponding effective potential can be found with aid of (\ref{veff}) 
\begin{equation}
\label{veff1}
V_{\textrm{eff}}(x)=\frac{1}{4}+\frac{1}{36}(1-\alpha^2)^2(1-2\alpha^2)^2(3-4\alpha^2)l^2-\frac{1}{2(1-\alpha^2)x}+\frac{l^2}{4x^2}-\frac{l^2}{2(1-\alpha^2)x^3}+\frac{1}{36}(1-\alpha^2)^2(1-2\alpha^2)^2(3-4\alpha^2)x^2\ .
\end{equation}
Here we followed the same receipt performed in the previous section in order to examine the nature of the motion of possible test particles under the influence of the effective potential above. As the system under analysis includes a global monopole, we keep using as reference the same usual expression for the angular momentum, namely $L_{\textrm{\tiny ISCO}}=2\sqrt{3}GM/(1-8 \pi G \eta^2)$. Note that for this case we should have $\alpha^2$ as the only free parameter of the model, due to the constraint (\ref{psi0}). Taking it into account we looked for possible values of $\alpha^2$ capable to provide stable and circular motion, however we could not find any interesting result in this regard. We noticed that the necessary conditions mentioned before that has to be respected by a given effective potential in order to enable stable and circular orbits are not observed in the present case. We found that real and positive roots for $V'_{\textrm{eff}}(x)=0$ are only possible for the range of values $\alpha^2 \geq 0.75$ (given that $\alpha^2<1$), and all these values provide $V''_{\textrm{eff}}(x)>0$ indicating the inviability of such a kind of dynamics for test particles moving in this spacetime.   

\section{The light deflection}
In this section we investigate the effects arising from both the modified gravity and the influence of the global monopole on the phenomenon of the gravitational bending of light. Since in the Einstein gravity such effect is usually investigated considering a weak field limit, it is convenient to restrict our analysis of the light deviation for the metric (\ref{solB}). 

As it is well known, the equation describing the geodesic paths of the photons through a given static and spherically symmetric spacetime is obtained by taking $\epsilon=0$ in (\ref{Lg}), which will give us the orbit equation below
\begin{equation}
\label{geodL} 
\frac{1}{r^4}\left(\frac{dr}{d\varphi}\right)^2+\frac{B(r)}{Y_0 r^2}=\frac{1}{b^2},
\end{equation}
where $b\equiv L/E$ is the impact parameter of the light ray. For the case under consideration, $B(r)$ is given by (\ref{solB}) whereas $Y_0=1$. This equation allows us to obtain a typical quantity of the light bending physics, which is the the closest approach distance of the light ray with respect to the central mass, denoted by $r_0$. In practice, this parameter means the value of $r(\varphi)$ where the light path experiences a turning point and the radial motion reaches a minimum value, so $\frac{dr}{d\varphi}=0$. This condition when used in (\ref{geodL}) provides the algebraic equation below for $r_0$
\begin{equation}
\label{r0} 
r_0^{3}+\psi_0 b^2 r_0 -b^2(1-\alpha^2)r_0+2GMb^2=0.
\end{equation}
Let us recall that $\alpha^2\equiv 8 \pi G \eta^2$, as previously introduced in (\ref{solX}). The solution for this cubic equation can be written as follows
\begin{equation}
\label{r0sol} 
r_0=\frac{2b}{3}\sqrt{\psi_0^2b^2+3(1-\alpha^2)}\textrm{cos}\left\{\frac{1}{3}\textrm{cos}^{-1}\left[\frac{-54GM-\psi_0\left(2\psi_0^2b^2+9(1-\alpha^2)\right)b^2}{2b\left(\psi_0^2b^2+3(1-\alpha^2)\right)^{3/2}}\right]\right\}-\frac{\psi_0b^2}{2}. 
\end{equation}
In Appendix \ref{A} we present the detailed derivation of (\ref{r0sol}). Notice that when one sets $\alpha^2=0$ and $\psi_0=0$ in the expression (\ref{r0sol}), it then becomes 
\begin{equation}
\label{r0solRG}
r_0=\frac{2b}{\sqrt{3}}\textrm{cos}\left[\frac{1}{3}\textrm{cos}^{-1}\left(\frac{-3\sqrt{3}GM}{b}\right)\right],
\end{equation}
which is its GR counterpart, as can be checked in \cite{wald}. If we keep up to first order contributions of $GM$ and $\psi_0$ in (\ref{r0sol}), we find the linearized form for the closest approach distance:
\begin{equation}
\label{1str0}
r_0\simeq b\sqrt{1-\alpha^2}-\frac{GM}{1-\alpha^2}-\frac{\psi_0b^2}{2},
\end{equation}
from which we easily get the following relation 
\begin{equation}
\label{1str01}
\frac{1}{r_0}\simeq\frac{1}{b\sqrt{1-\alpha^2}}+\frac{GM}{b^2\left(1-\alpha^2\right)^2}+\frac{\psi_0}{2\left(1-\alpha^2\right)},
\end{equation}
which is going to be useful to us later on.

It is usual to define the radial coordinate as $r=u^{-1}$, so that (\ref{geodL}) turns out to be
\begin{equation}
\label{geodL1} 
\left(\frac{du}{d\varphi}\right)^2+\frac{{\cal B}(u)u^2}{Y_0}=\frac{1}{b^2}, 
\end{equation} 
where ${\cal B}(u)=B(u(r))$ denotes the reparametrization of $B$ in terms of $u$. For the background metric (\ref{solB}) this means
\begin{equation}
\label{solB1}
{\cal B}(u)=1-\alpha^2-2GMu-\frac{\psi_0}{u},
\end{equation}
along with $Y_0=1$. This allows us to rewrite (\ref{geodL1}) as
\begin{equation}
\label{geodL2} 
\left(\frac{du}{d\varphi}\right)^2+(1-\alpha^2)u^2-2GMu^3-\psi_0 u=\frac{1}{b^2}. 
\end{equation}
Taking the derivative of (\ref{geodL2}) with respect to $\varphi$ we obtain
\begin{equation}
\label{geodL3}
\frac{d^2 u}{d \varphi^2}+(1-\alpha^2)u-3GM u^2-\frac{\psi_0}{2}=0,
\end{equation}
which is the nonlinear orbit equation corrected by the $f(R)$ parameter, $\psi_0$.

A crucial point in the present analysis is the non-asymptotically flat nature of the spacetime (\ref{solB}), which has decisive consequences on the calculation of the deflection angle. In an asymptotically flat geometry, like standard Schwarzschild case, one assumes that the light ray comes from the infinity where its path correspond to a straight line. As it approaches a spherical distribution of mass its trajectory gradually departs from the straight line getting closer and closer to the central mass, reaches a distance of closest approach with respect to it and then is bent by an angle $\delta$ proceeding towards the observer which is assumed to be located at infinity. For a Schwarzschild solution (to the leading order in $GM$) the magnitude of this deviation is given by $\delta =\frac{4GM}{c^2b}$.
However, for a non-asymptotically flat spacetime it does not make any sense to consider a light ray emitted at infinity (neither detected by an observer at infinity), due to the existence of a horizon at a given finite radial distance far from the spherical mass. This feature suggests that the procedure to compute the bending angle for a non-asymptotically flat background should be somehow different from the standard method used for the Schwarzschild spacetime.

The typical case where such discussion is usually raised is in the context of a SdS metric. During a long time, many authors claimed that the cosmological constant should not give any contribution to the light deflection, as it is absent in the second order differential equation for the orbit \cite{islam,freire,kerr,kunz,finelli,sereno}.
On the other hand, in \cite{rindler} the authors demonstrated that a contribution of $\Lambda$ to the bending in fact emerge from the structure of the SdS spacetime itself and introduced a method for calculating explicitly such effect for a given static and radial metric, considering a light ray emitted from a finite arbitrary position $P(r(\varphi),\varphi)$. Besides, they also emphasized that a non-zero effect of the cosmological constant should arise, since $\Lambda$ does appear in the first integral of the second order orbit equation, namely (\ref{geodL}), and one expects that the corresponding solution $u(\varphi)$ obeys the both. So, (\ref{geodL}) would work as a complementary equation necessary to fix the integration constants of the solution, endowing the general solution $u(\varphi)$ necessarily with a $\Lambda-$dependence. 
Since then, other authors reexamined the problem, seeking to fill possible gaps remaining in the original Rindler's approach \cite{bhadra,arakida,pacheco}. For instance, \cite{bhadra} and \cite{pacheco} investigate the influence on the deflection angle of the angular position both of the source and the observer, showing that these two variables should play an important role in the understanding of the bending of light in SdS spacetime. Likewise, finite-distance corrections on the light deflection was also explored in \cite{ishihara}. In \cite{oliver} the author looked for possible effects of the background expansion on the bending angle, however he did not find any contribution. For sake of simplicity and bearing in mind that this issue keeps being matter of vivid debate in the literature, for this moment we will restrict ourselves to the Rindler-Ishak formalism in order to assess the impact of the contributions of both the modified gravity and the global monopole. We will postpone a deeper and more detailed and analysis to a future opportunity.

Throughout this paper we are considering a Schwarzschild black hole which swallowed a global monopole and incorporated its charge. However, notice that even without such interaction with the black hole, the defect is able to affect the trajectory of light particles moving nearby due to the solid deficit angle that appears in its surroundings. So, it is convenient to provide a definition for the angle $\varphi$ that accounts for the residual influence of the solid deficit angle even if $M=0$ (as well as $\psi_0=0$). In this vein, from now on we will consider the following change of variable $\sqrt{1-\alpha^2}\varphi \longrightarrow \bar{\varphi}$ and hence $u(\varphi(\bar{\varphi}))=\bar{u}(\bar{\varphi})$, which makes (\ref{geodL2}) to be rewritten as 
\begin{equation}
\label{geodL4}
\left(\frac{d\bar{u}}{d\bar{\varphi}}\right)^2+\bar{u}^2-\frac{2GM\bar{u}^3}{(1-\alpha^2)}-\frac{\psi_0 \bar{u}}{(1-\alpha^2)}=\frac{1}{b^2(1-\alpha^2)},
\end{equation}
whereas (\ref{geodL3}) now is
\begin{equation}
\label{geodL5}
\frac{d^2 \bar{u}}{d \bar{\varphi}^2}+\bar{u}-\frac{3GM \bar{u}^2}{(1-\alpha^2)}-\frac{\psi_0}{2(1-\alpha^2)}=0.
\end{equation}
One possible way to solve (\ref{geodL4}) is resorting to a perturbative method, in which the function $\bar{u}(\bar{\varphi})$ is split into the different perturbative orders. Here we will consider up to first order effects on the bending of light, which implies to adopt the following decomposition 
\begin{equation}
\label{u01}
\bar{u}=\bar{u}_0+\bar{u}_1.
\end{equation}
For our purposes both $GM$ and $\psi_0$ will be considered as first order quantities. Substituting (\ref{u01}) in (\ref{geodL4}) we are left with the two equations below
\begin{equation}
\label{u0}
\frac{d^2 \bar{u}_0}{d \bar{\varphi}^2}+\bar{u}_0=0
\end{equation}
and
\begin{equation}
\label{u1}
\frac{d^2 \bar{u}_1}{d \bar{\varphi}^2}+\bar{u}_1=\frac{3GM \bar{u}_0^2}{(1-\alpha^2)}+\frac{\psi_0}{2(1-\alpha^2)},
\end{equation}
at zeroth and first orders, whose solutions are given by
\begin{equation}
\label{solu0} 
 \bar{u}_0=\frac{\textrm{sin}\bar{\varphi}}{R}
\end{equation}
and
\begin{equation}
\label{solu1} 
 \bar{u}_1=\frac{3GM}{2R^2(1-\alpha^2)}\left(1+\frac{1}{3}\textrm{cos}2\bar{\varphi}\right)+\frac{\psi_0}{2(1-\alpha^2)},
\end{equation}
respectively. Hence, using (\ref{u01}) we write the linearized solution for (\ref{u1}) 
\begin{equation}
\label{solu}
\bar{u}=\frac{\textrm{sin}\bar{\varphi}}{R}+\frac{3GM}{2R^2(1-\alpha^2)}\left(1+\frac{1}{3}\textrm{cos}2\bar{\varphi}\right)+\frac{\psi_0}{2(1-\alpha^2)}.
\end{equation}
In the solution above the integration constants are chosen with the aid of the initial conditions $\bar{u}(\pi/2)=1/r_0$ and $\frac{d\bar{u}(\bar{\varphi})}{d \bar{\varphi}}|_{\bar{\varphi}=\pi/2}$, which means to assume a symmetric scheme where the light ray reaches its turning point in the middle of its full path, whose corresponding angular position is $\bar{\varphi}=\pi/2$, comprised between the light source and the observer. The first condition tell us that in order to have (\ref{solu}) consistent with (\ref{1str01}) it is necessary to fix the remaining integration constant $R$ as $R=\sqrt{1-\alpha^2}b$.
 
Following the Rindler-Ishak procedure, we define $\Psi$ as the angle between the radial direction and the light trajectory at a given point $P(r,r(\varphi))$. It is easy to check that $\Psi$ is related to the angular position $\varphi$ through the equation below
\begin{equation}
\label{rind}
\textrm{tan} \Psi=r\sqrt{B(r)}\left|\frac{dr}{d\varphi}\right|^{-1}.
\end{equation}
Given a angular position $\varphi$, the corresponding radial one $r(\varphi)$ is immediately found from the solution of the orbit equation. For the Schwarzschild case, the desired bending angle $\delta$ is just the double of the magnitude of $\Psi$. However, as pointed out in \cite{arakida}, this relation is based on fundamental properties of the euclidean geometry, and can be only applied for asymptotically flat spaces and very far from the central mass. So, we have to bear in mind that it is not correct to obtain the bending angle merely by $\delta=2|\Psi|$ in a non-asymptotically flat background, as is usually done in the Schwarzschild context. 
Nevertheless, $\Psi$ can give us a primary idea about the influence on the light deflection of non-asymptotically flat corrections on the Schwarzschild metric.

It is convenient to make a change of variable in (\ref{rind}) and rewrite it in terms of $\bar{u}$ and $\bar{\varphi}$. Furthermore, assuming  a small enough $\Psi$, the following approximation holds 
\begin{equation}
\label{rind1} 
\textrm{tan} \Psi \simeq \Psi \simeq \sqrt{{\cal B}(\bar{u})}(1-\alpha^2)^{-1/2}\bar{u}\left|\frac{d\bar{u}}{d\bar{\varphi}}\right|^{-1}.
\end{equation}
Now, let us consider the specific example $\bar{\varphi}=0$. For this case, using (\ref{rind1}) we obtain the following bending angle
\begin{equation}
\label{bend}
\Psi\simeq \frac{2GM}{b\left(1-\alpha^2\right)^{3/2}}\sqrt{1-\frac{\psi_0^2b^4(1-\alpha^2)}{(4GM)^2}}.
\end{equation}
Taking $\psi_0$ and $\alpha^2$ in the result above, we have the standard bending angle $\delta=2\left|\Psi\right|=\frac{4GM}{b}$ (assuming $c=1$), where one usually considers $\varphi \sim 0$ and $r \rightarrow \infty$ ($u \rightarrow 0$). This result indicates that the modified gravity contributes to the decreasing of $\Psi$. On the other hand, if $\alpha^2=0$ and the departure from GR is not significant when comparing to $GM$, so that $\delta\simeq 2\left|\Psi\right|$, the reduction of the Schwarzschild's bending angle due to the modified of gravity corrections shall be negligible.


\section{Concluding remarks}
In this paper we revisit and provide some contributions to the study of the $f(R)$ global monopole. We considered the hypothetical case where a global monopole were swallowed by a Schwarzschild black hole, within a $f(R)$ gravity framework. For such system we derived the field equations in the metric formalism, according to the Multamaki-Vilja \cite{Mut} method and obtained an exact solution for the problem. We showed how to obtain some particular cases from this general solution. We also studied the asymptotic behavior of the Higgs field very far from the monopole's core, demonstrating the explicit dependence on the background geometry for a wide class of static and radial metrics. This result extends other ones obtained previously in the context of a dS/AdS spacetime \cite{bertrand}. In order to better understand the gravitational effects of the $f(R)$ global monopole, we studied the motion of test particles in its spacetime, analyzing the conditions for obtaining stable and circular orbits. In this case, we found that stable orbits keeps being possible, however their positions are slightly shifted far from the black hole as the modification of the gravity is increased. Furthermore, we noticed that unstable orbits turned out to be allowed as well, differently from what happens in GR.

We finished our analysis by studying the bending of light for a $f(R)$ global monopole, focusing on the metric (\ref{solB}). Since for this case the spacetime is not asymptotically flat, an alternative procedure for the calculation of the bending has to be employed. In line with Rindler-Ishak formalism \cite{rindler}, we computed the angle $\Psi$ between the light trajectory and its radial position using $\varphi=0$ for the angular position of the light ray. In the standard Schwarzschild case $\Psi$ is just half the bending angle. We found that the modification of the gravity $\psi_0$ contributes for the decreasing of this angle. Anyway, if $\psi_0$ is small enough so that $\Psi$ can be considered as the half of the defletion angle, the deviation from the standard GR situation shall be negligible. We are aware of the importance of extending our analysis and we totally agree that in a more complete study, based on a non-asymptotically flat spacetime, the positions of both the source and the observer should necessarily come into play. However, we just wanted to give an illustration about some imediate gravitational consequences of the $f(R)$ global monopole, so we leave a deeper investigation in this respect for a future work. 


\noindent
{\bf Acknowledgement:} The authors are grateful to CNPq (Brazil) and FAPES (Brazil) for partial financial support. ERBM has been partially supported by CNPq through the project No.
313137/2014-5. We also thank J. Freitas Pacheco, O. Piattella and D. C. Rodrigues for much appreciated discussions.

\appendix

\section{Solving the cubic equation for $r_0$}
\label{A}
For sake of simplicity, let us rewrite (\ref{r0}) in the form
\begin{equation}
\label{r0A}
x^3+dx^2-cx+a=0,
\end{equation}
where we clearly have $x=r_0$, $d=\psi_0b^2$, $c=(1-\alpha^2)b^2$ and $a=2GMb^2$. We are interested in expressing $x$ in terms of a new variable $w$ in the following way
\begin{equation}
\label{xAB}
x=A \cos{w}+B,
\end{equation}
where $A$ and $B$ are arbitrary constants to be fixed later in terms of the parameters $a$, $c$ and $d$. So, we shall look for $A$, $B$ and $w$ so that the relation (\ref{xAB}) is possible. When we substitute (\ref{xAB}) into (\ref{r0A}) we have
\begin{equation}
\label{algebr0}
(A\cos{w}+B)^3+d (A\cos{w}+B)^2-c (A\cos{w}+B)+a=0.
\end{equation}
Now let us expand (\ref{algebr0}) by collecting terms in powers of $\cos{w}$ as follows 
\begin{equation}
\label{eqA0} 
A^3 \cos^3 {w} + A^2(3B+d)\cos^2{w}+A(3B^2+2Bd-cA)\cos{w}+a+B^3-cB+B^2d=0.
\end{equation}
Then, by using the trigonometric identity $4\cos^3{w}=\cos{w}+3\cos{w}$, we can eliminate $\cos^3{w}$ in (\ref{eqA0}), so
\begin{eqnarray}
\label{eqA1}
&&A^{3}\left(\frac{\cos{3w}}{4}+\frac{3\cos{w}}{4}\right)+A^2(3B+d)\cos^2{w}+A(3B^2+2Bd-cA)\cos{w}+a+B^3-cB+B^2d=0, \nonumber\\
&&\frac{A^{3}}{4}\cos{3w}+A^2(3B+d)\cos^2{w}+A\left(\frac{3A^2}{4}+3B^2+2Bd-cA\right)\cos{w}+a+B^3-cB+B^2d=0.
\end{eqnarray}
For our purposes, in order to better compare with the general relativistic equation (\ref{r0solRG}), we want to express (\ref{eqA1}) as 
$\cos{3w}\propto \textrm{const.}$ This requires the vanishing of the coefficients both of $\cos^2{w}$ and $\cos{w}$ in (\ref{eqA1}), resulting in the following system of equations: 
\begin{equation}
\label{sAB}
\begin{cases}
3B+d=0,\\
\frac{3A^2}{4}+3B^2+2Bd-cA=0.
\end{cases}
\end{equation}
Notice that the only unknowns in the equation above are $A$ and $B$, since $c$ and $d$ were introduced as redefinitions for the quantities $(1-\alpha^2)b^2$ and $\psi_0b^2$, respectively. So, the system above indeed admits a single solution ${\cal S}_0=\{A,B\}$. From (\ref{sAB}) it is easy to find
\begin{equation}
\label{sB}
B=-d/3
\end{equation}
and
\begin{equation}
\label{sA}
A=\frac{2}{3}\sqrt{d^2+3c}.
\end{equation}
With (\ref{sB}) and (\ref{sA}) the equation (\ref{eqA1}) reduces to
\begin{eqnarray}
\label{eqA2} 
\cos{3w}=-\left[\frac{27a+d\left(2d^2+9c\right)}{2(d^2+3c)^{3/2}}\right],
\end{eqnarray}
which gives us
\begin{equation}
\label{eqA3}
w=\frac{1}{3}\cos^{-1}\left[\frac{-27a-d\left(2d^2+9c\right)}{2(d^2+3c)^{3/2}}\right].
\end{equation}
Now let us take $\cos$ on the both sides of (\ref{eqA3})
\begin{equation}
\label{eqA4}
\cos{w}=\cos\left\{\frac{1}{3}\cos^{-1}\left[\frac{-27a-d\left(2d^2+9c\right)}{2(d^2+3c)^{3/2}}\right]\right\}.
\end{equation}
Substituting (\ref{sB}), (\ref{sA}) and (\ref{eqA4}) into (\ref{xAB}) we arrive at the desired expression for $x$ 
\begin{eqnarray}
\label{xAB1}
&&x=A\cos\left\{\frac{1}{3}\cos^{-1}\left[\frac{-27a-d\left(2d^2+9c\right)}{2(d^2+3c)^{3/2}}\right]\right\}+B\nonumber\\
&&=\left(\frac{2}{3}\sqrt{d^2+3c}\right)\cos\left\{\frac{1}{3}\cos^{-1}\left[\frac{-27a-d\left(2d^2+9c\right)}{2(d^2+3c)^{3/2}}\right]\right\}-\frac{d}{3}.
\end{eqnarray}
Now we are ready to recover the original values of $a$, $c$, $d$ and $x$, as it has been defined in the beginning of this Appendix. This will allows us to write (\ref{xAB1}) as 
\begin{equation}
r_0=\frac{2b}{3}\sqrt{\psi_0^2b^2+3(1-\alpha^2)}\textrm{cos}\left\{\frac{1}{3}\textrm{cos}^{-1}\left[\frac{-54GM-\psi_0\left(2\psi_0^2b^2+9(1-\alpha^2)\right)b^2}{2b\left(\psi_0^2b^2+3(1-\alpha^2)\right)^{3/2}}\right]\right\}-\frac{\psi_0b^2}{2}, 
\end{equation}
which is exactly the equation (\ref{r0sol}).



\begin{thebibliography}{100}
\bibitem{kibble} T. W. B. Kibble, J. Phys. A {\bf 9}, 1387 (1976).
\bibitem{vilenkin} A. Vilenkin and E. P. Shellard, {\it Cosmic String and Other Topological Defects} (Cambridge University Press, Cambridge, England, 1994).
\bibitem{sokolov} D. D. Sokolov and A. A. Starobinsky, Dokl. Akad. Nauk USSR. {\bf 234}, 1043 (1977) [Sov.Phys. - Doklady {\bf 22}, 312 (1977)].
\bibitem{barriola} M. Barriola and A. Vilenkin, Phys. Rev. Lett. {\bf 63}, 341, (1989).
\bibitem{lousto} D. Hahari and C. Loust\`o, Phys. Rev. D {\bf 42} 2626, (1990).
\bibitem{staro} A. A. Starobinsky, Phys. Lett. B {\bf 91} 99, (1980).
\bibitem{capo} S. Capozziello, A. Stabile, A. Troisi, Phys. Lett. B {\bf 686}, 79 (2010).
\bibitem{soti} Thomas P. Sotiriou, Class.Quant.Grav. {\bf 23}, 5117-5128, (2006).
\bibitem{sotiriou} Thomas P. Sotiriou and Valerio Faraoni, Rev. Mod. Phys. {\bf 82}, 451 (2010).
\bibitem{odin} S. Nojiri and S. D. Odintsov, Phys. Rept. {\bf 505}, 59-144  (2011). [arXiv:1011.0544]
\bibitem{odin1} S. Nojiri, S.D. Odintsov and V.K. Oikonomou, arXiv:1705.11098.
\bibitem{romero} A. Barros and C. Romero, Phys. Rev D {\bf 56}, 6688, (1997).
\bibitem{carames1} T. R. P. Caram\^es, E. R. Bezerra de Mello, M. E. X. Guimar\~aes, Int. J. Mod. Phys. D: Conference Series, v. 03, p. 446-454, (2011).  
\bibitem{carames2} T. R. P. Caram\^es, E. R. Bezerra de Mello, M. E. X. Guimar\~aes, Mod. Phys. Lett. A, v. 27, No. 30 1250177 (2011).
\bibitem{man} Jingyun Man and Hongbo Cheng, Phys. Rev. D {\bf 87}, 044002 (2013).
\bibitem{cheng} Jingyun Man and Hongbo Cheng, Phys. Rev. D {\bf 92}, 024004 (2015).
\bibitem{jpmg} J. P. Morais da Gra\c{c}a and V. B. Bezerra, Mod. Phys. Lett. A, v. 27, 1250178 (2012).
\bibitem{valdir} J. P. M. Gra\c{c}a, H. S. Vieira, V. B. Bezerra, Gen. Rel. Grav. 48, No. 4, 38 (2016).
\bibitem{melis} Melis U. Dogru and Dogukan Taser, Mod. Phys. Lett. A, v. 30, No. 40 1550217 (2015).
\bibitem{Mut} T. Multamaki and I. Vilja, Phys. Rev. D {\bf 74}, 064022 (2006).
\bibitem{bertrand} B. Bertrand, Y. Brihaye and B. Hartmann, Class. Quantum Grav. {\bf 20}, 4495-4502 (2003).
\bibitem{mann} P. D. Mannheim and D. Kazanas, Astrophys. J. 342, 635 (1989).
\bibitem{kaplan} S. A. Kaplan, Zh. Eksp. Teor. Fiz. 19, 951 (1949).
\bibitem{landau} L. D. Landau and E. M. Lifshitz, {\it The Classical Theory of Fields} (Pergamon, Oxford, 1993).
\bibitem{dadhich} N. Dadhich, K. Narayan and U. A. Yajnik, Pramana- Journal of Physics {\bf 50}: 302-314 (1998).
\bibitem{wald} R. M. Wald, {\it General Relativity} (The University of Chicago Press, Chicago and London, 1984).
\bibitem{islam} N. J. Islam, Phys. Lett. {\bf 97A}, 239 (1983)
\bibitem{freire} W. H. C. Freire, V. B. Bezerra and J. A. S. Lima, Gen. Relativ. Gravit. {\bf 33}, 1407 (2001).
\bibitem{kerr} A. W. Kerr, J. C. Hauck and B. Mashhoon, Class. Quantum Grav. {\bf 20}, 2727 (2003).
\bibitem{kunz} V. Kagramanova, J. Kunz, and C. Lammerzahl, Phys. Lett. {\bf B634}, 465 (2006).
\bibitem{finelli} F. Finelli, M. Galaverni, and A. Gruppuso, Phys. Rev. D {\bf 75}, 043003 (2007).
\bibitem{sereno} M. Sereno and Ph. Jetzer, Phys. Rev. D {\bf 73}, 063004 (2006).
\bibitem{rindler} W. Rindler and M. Ishak, Phys. Rev. D {\bf 76}, 043006 (2007). 
\bibitem{bhadra} A. Bhadra, S. Biswas and K, Sarkar, Phys. Rev. D {\bf 82}, 063003 (2010).
\bibitem{arakida} H. Arakida and M. Kasai, Phys. Rev. D {\bf 85}, 023006 (2012).
\bibitem{pacheco} T. Biressa and J. A. de Freitas Pacheco, Gen. Relativ. Gravit. {\bf 43}, 2649 (2011). 
\bibitem{ishihara} A. Ishihara, Y. Suzuki, T. Ono, T. Kitamura and H. Asada, Phys. Rev. D {\bf 94}, 084015 (2016). 
\bibitem{oliver} O. Piattella, Phys. Rev. D {\bf 93}, 024020 (2016).
\end{thebibliography}
\end{document}